\documentclass[%
reprint,
superscriptaddress,
amsmath,amssymb,
aps,
prl,
floatfix,
]{revtex4-1}

\usepackage{color}
\usepackage{bbold}
\usepackage{graphicx}
\usepackage{dcolumn}
\usepackage{bm}
\usepackage{hyperref}
\usepackage[normalem]{ulem}

\hypersetup{
    colorlinks=true,
    linkcolor=blue,
    citecolor=blue,
    filecolor=black,
    urlcolor=blue,
}

\newcommand{\order}[1]{\ensuremath{\mathcal{O}(#1)}}

\def\vev#1{ \left\langle #1 \right \rangle }

\def\Nf{N_f}

\begin{document}

\title{Gravitational wave consistency relations for multifield inflation}

\author{Layne C. Price}
  \email{lpri691@aucklanduni.ac.nz}
  \affiliation{Department of Physics, University of Auckland, Private Bag 92019,  Auckland, New Zealand}

\author{Hiranya V. Peiris}
  \email{h.peiris@ucl.ac.uk}
  \affiliation{Department of Physics and Astronomy, University College London, London WC1E 6BT, U.K.}

\author{Jonathan Frazer}
  \email{j.frazer@ucl.ac.uk}
  \affiliation{Department of Theoretical Physics, University of the Basque Country UPV/EHU, 48040 Bilbao, Spain
}

\author{Richard Easther}
  \email{r.easther@auckland.ac.nz}
  \affiliation{Department of Physics, University of Auckland, Private Bag 92019,  Auckland, New Zealand}

\date{\today}

\begin{abstract}
  We study the tensor spectral index $n_t$ and the tensor-to-scalar ratio $r$ in the simplest multifield extension to single-field, slow-roll inflation models.  We show that multifield models with potentials $V \sim \sum_i \lambda_i |\phi_i|^p$ have different predictions for $n_t/r$ than single-field models, even when all the couplings are equal $\lambda_i=\lambda_j$, due to the probabilistic nature of the fields' initial values.  We analyze well-motivated prior probabilities for the $\lambda_i$ and initial conditions to make detailed predictions for the marginalized probability distribution of $n_t/r$.  With $\mathcal O(100)$ fields and $p>3/4$, we find that $n_t/r$ differs from the single-field result of $n_t/r=-1/8$ at the 5$\sigma$ level.  This gives a novel and testable prediction for the simplest multifield inflation models.
\end{abstract}

\maketitle

A  cosmological gravitational wave background (CGWB) is a compelling signature of inflation, which is already supported by the highly Gaussian  primordial perturbations \cite{Ade:2013ydc,Leistedt:2014zqa} and their broken scale invariance, now detected at $5\sigma$  significance \cite{Ade:2013zuv, Ade:2013uln}. A  large-amplitude CGWB   provides fundamentally new tests of  single-field slow-roll (SFSR) inflation via the \emph{consistency relation} \cite{Copeland:1993ie}  $n_t/r = -1/8$, which relates the tensor spectral index $n_t$ to the ratio of the tensor and scalar perturbation amplitudes, $r$.

While there has been dramatic progress towards the direct detection of a CGWB through the $B$-mode polarization in the cosmic microwave background (CMB) \cite{Ade:2014xna}, measuring $n_t$ is challenging with current technologies \cite{Verde:2005ff,Dodelson:2014exa,Caligiuri:2014sla}.
However, for $r \gtrsim 0.1$ this will be feasible with the next generation of space-based~\cite{Baumann:2008aq,Andre:2013afa}, ground-based~\cite{Keckarray,BICEP3,PolarBear2,SPT3G}, and balloon-borne~\cite{EBEX,Spider} experiments, while future 21 cm projects~\cite{Masui:2010cz,Book:2011dz}  could also detect lensing by a CGWB and direct detection experiments \cite{Smith:2006xf,Boyle:2014} would test the consistency condition using the lever arm between  CMB and solar system scales to far greater accuracy with $r \gtrsim \order{10^{-3}}$.

The simplest inflationary scenarios that yield an easily detectable CGWB are  \emph{monomial} models with the inflationary potential  $V \sim |\phi|^p$, which have $0.05 \lesssim r \lesssim 0.30$ for $2/3 \lesssim p \lesssim 4$. Single field  models are {\em simple\/} but not necessarily {\em natural\/}, as many high energy theories  yield large numbers of scalar degrees of freedom~\cite{Grana:2005jc,Douglas:2006es,Denef:2007pq,Denef:2008wq}.
For multifield models the consistency relation is reduced to an inequality, $n_t/r\le-1/8$.  While $r$ and $n_t$ are correlated for $\Nf=2$~\cite{Bartolo:2001rt,Wands:2002bn}, there is no known relationship between $r$ and $n_t$ when $\Nf$ is large.

In this {\em Letter\/}, we derive a robust prediction for  $n_t/r$ for  \emph{$\Nf$-monomial} models, with potential
\begin{equation}
  V = \frac{1}{p} \sum_{i} \lambda_i |\phi_i|^p,
  \label{eqn:V}
\end{equation}
where $\lambda_i$ are real, positive  constants and  summations run over the number of fields, $\Nf$.
 Eq.~\eqref{eqn:V} arises naturally in many high energy theories~\cite{Liddle:1998jc,Kanti:1999vt,Kanti:1999ie,Kaloper:1999gm,Kanti:1999ie,Easther:2005zr,Dimopoulos:2005ac,Kim:2006ys,Kim:2007bc} and  is a simple, intuitive generalization of the chaotic SFSR models.

 We treat the $\lambda_i$ and the values of $\phi_i$ at a fixed number of $e$-folds before the end of inflation as independent random variables. When $\Nf \to \infty$, the central limit theorem ensures  that $n_t/r$ is a Gaussian random variable.  Critically,   $\vev{n_t/r}$ does not reduce to the single-field limit if the couplings are identical unless the field values $\phi_{i,*}$ when the pivot scale $k_*$ leaves the horizon are also fixed, except for the special case $p=2$. The expected value of $n_t/r$  depends only on two moments of the distributions of the $\lambda_i$ and $\phi_i$, and is independent of $\Nf$.  The variance in $n_t/r$ is $s^2_{n_t/r} \sim 1/\Nf$ (for $p>3/4$),  giving a sharp, generic prediction for the consistency relation in the many-field limit.
 This provides a direct test for distinguishing between $\Nf$-monomial models and their single-field analogues.

{\bf Model---}
In some cases the $\lambda_i$ in Eq.~\eqref{eqn:V} might be derivable from fundamental theory, but in general we are ignorant of their values, so we treat these terms as independent random variables (RVs) with a prior probability $P(\lambda)$.
Similarly, we do not know the fields' initial conditions, so we also treat these as identically distributed, but possibly correlated, RVs with a prior probability $P(\phi_0)$.
We then marginalize over the  $P(\lambda)$ and $P(\phi_0)$ to produce a probability distribution for $n_t/r$.
Since a change of variables $\phi_i \to \tilde \phi_i(\phi_j,\lambda_j)$ will mix the $\lambda_i$ and $\phi_i$, it is clear that there is no \emph{a priori} difference between these two types of parameters, motivating our statistical approach.

The simplest choice for $P(\phi_0)$ is a uniform distribution of $\phi_{i,*}$ defined when the pivot scale $k_*$ leaves the horizon $N_*$ $e$-folds before the end of inflation.  This choice contains the least Shannon information about the fields' initial states and ensures that most of the fields are dynamically relevant.
Further, this $P(\phi_0)$ and others were extensively studied in Ref.~\cite{Easther:2013rva}, where it was shown that the initial conditions only weakly affect the predicted density spectra.
The likely values of $n_s$ and $r$ for a related class of multifield monodromy models was derived in Ref.~\cite{Wenren:2014cga}, finding $0.955 \lesssim n_s \lesssim 0.975$.  Furthermore, $r=4p/N_*$, and the non-Gaussianity is small.

{\bf $\delta N$ formalism---}
The potential in Eq.~\eqref{eqn:V} is sum-separable and, assuming slow-roll,  $N_*$ is~\cite{Vernizzi:2006ve,Battefeld:2006sz}
\begin{equation}
  N_* = - \int_*^c \sum_i \frac{V_i}{V_i'} d \phi_i,
  \label{eqn:XXX}
\end{equation}
where $V_i' = \lambda_i |\phi_i|^{p-1}$ and $\phi_{i,*}$ and $\phi_{i,c}$ denote field values at horizon crossing and the end of inflation, respectively.  For $\Nf$-monomial inflation
\begin{equation}
  N_* = \frac{1}{2p} \sum_{i} \left[ \phi_{i,*}^2 - \phi_{i,c}^2 \right].
  \label{eqn:N_hca}
\end{equation}

The $\delta N$ formalism  relates the field perturbations at horizon crossing to the gauge-invariant curvature perturbation $\zeta$ on constant density hypersurfaces via
\begin{equation}
  \zeta \approx \sum_i N_{*,i} \delta \phi_{i,*},
  \label{eqn:deltaN}
\end{equation}
where $N_{*,i} \equiv \partial N_*/\partial  \phi_{i,*}$.  If  the field perturbations are well-approximated by a free field theory with  power spectrum $\mathcal P_{\delta \phi}^{ij} = ( H_*/2 \pi )^2 \delta^{ij}$  at horizon crossing,  the tensor-to-scalar ratio is
\begin{equation}
  r = \frac{8}{\sum_i N_{*,i} N_{*,i}}.
  \label{eqn:XXX}
\end{equation}
To first-order in slow-roll $n_t = -2 \epsilon$, where
\begin{equation}
  \epsilon = \frac{1}{2} \sum_i \left[ \frac{V_i^\prime}{V} \right]^2 .
  \label{eqn:eps}
\end{equation}

For $\Nf$-monomial models, the field values $\phi_{i,c}$ at the end of inflation can typically be neglected.  This \emph{horizon crossing approximation} (HCA) ({\it e.g.,} Refs.~\citep{Vernizzi:2006ve,Kim:2006te}) is a simplification of the  $\delta N$ formalism that incorporates the super-horizon evolution of $\zeta$, but ignores quantities contributing to $N_*$ from the end-of-inflation surface.  Setting $\phi_{i,c} \to 0$ in Eq.~\eqref{eqn:N_hca}, we find
\begin{eqnarray}
  \frac{n_t}{r} &=& - \frac{1}{4p^2}\epsilon \sum_i \phi_{i,*}^2 \, ,
  \label{eqn:ntr_gen}
\end{eqnarray}
where we restrict our attention to cases that are slowly rolling at horizon crossing.  Requiring $\epsilon \lesssim 0.1$ then sets the maximum deviation from the single-field result as
\begin{equation}
  -\left(\frac{N_*}{2p}\right) \times \mathcal O(10^{-1})  \;\lesssim \; \frac{n_t}{r} \; \le \; -\frac{1}{8}.
  \label{eqn:XXX}
\end{equation}

{\bf The many-field limit---}
We build the probability distribution for $n_t/r$ by marginalizing Eq.~\eqref{eqn:ntr_gen} over $P(\phi_0)$ and $P(\lambda)$, and use the central limit theorem (CLT) to take the large $\Nf$ limit,  $\Nf \to \infty$.  By Eq.~\eqref{eqn:N_hca} the HCA implies that $P(\phi_0)$ is a uniform distribution pulled back onto an $\Nf$-sphere in field-space with radius $\sqrt{2p N_*}$.  Since the multivariate normal distribution $\vec{x} \sim \mathcal N (0,\mathbb{1})$ is invariant under rotations of $\vec x$, we can sample this $\Nf$-sphere uniformly by defining
\begin{equation}
  \phi_{i,*} = \sqrt{\frac{2 p N_*}{\sum_j x_j^2}} \; x_i
  \quad \mathrm{for} \quad
  \vec x \sim \mathcal N(0,\mathbb{1}).
  \label{eqn:phi_sphere}
\end{equation}

Using Eq.~\eqref{eqn:phi_sphere}, the summations in Eqs.~\eqref{eqn:eps}~and~\eqref{eqn:ntr_gen} are
\begin{equation}
  \sum_i \lambda_i^n | \phi_{i,*}|^m = \sum_i \lambda_i^n \left[ \frac{2 p N_*}{\sum_j x_j^2} \right]^{\frac{m}{2}} |x_i|^m.
  \label{eqn:sums}
\end{equation}
As $\Nf \to \infty$ the CLT shows that the numerator is normally distributed with mean
\begin{equation}
  \mu_\mathrm{num} = \Nf \left( 2 p N_* \right)^{m/2} \vev{\lambda^n} \vev{|x|^m}
  \label{eqn:XXX}
\end{equation}
and standard deviation
\begin{equation}
  s_\mathrm{num} = \sqrt{ \Nf} \left( 2 p N_* \right)^{m/2} \sigma_{n,m},
  \label{eqn:XXX}
\end{equation}
where $\vev{.}$ indicates expected value and
\begin{equation}
  \sigma_{n,m}^2 \equiv \vev{\lambda^{2n}} \vev{|x|^{2m}} - \vev{\lambda^n}^2 \vev{|x|^m}^2,
  \label{eqn:XXX}
\end{equation}
which assumes that the $\lambda_i$ and $x_j$ are independent.  Finally, the denominator in Eq.~\eqref{eqn:phi_sphere} is drawn from the $\chi$-distribution, which  is closely approximated by $\mathcal N(\sqrt{\Nf},1/\sqrt{2})$ for $x_i \sim \mathcal N(0,1)$.

The numerator and  denominator in Eq.~\eqref{eqn:sums} are  correlated by the constraint in Eq.~\eqref{eqn:N_hca}.  For a given variance in $P(\lambda)$, the correlation $\gamma$ is maximized when  $m=2$ and $|\gamma| \to 1$ as the variance vanishes.  Since each  $\sum_i \lambda_i^n |\phi_{i,*}|^m$ is uniquely determined given $\vec{\lambda}$ and $\vec{\phi}_{*}$, we expect a strong correlation between the numerator and denominator in Eq.~\eqref{eqn:eps} for typical choices of $P(\lambda)$. This significantly reduces the  variance of $n_t/r$, and ensures a sharp prediction for its value.  We numerically calculate $\gamma$ after defining the priors on $\lambda$.

For any normally distributed variable $y \sim \mathcal N(\mu, \sigma)$
\begin{equation}
\vev{ | y|^m} = \frac{2^{\frac{m}{2}} \sigma^m}{\sqrt{\pi}}   \Gamma\left(\frac{1+m}{2} \right) F_{1,1} \left( \frac{-m}{2}; \frac{1}{2};  \frac{-\mu^2}{2 \sigma^2} \right),
  \label{eqn:XXX}
\end{equation}
for $m>-1$, and  $F_{1,1}$ is the confluent hypergeometric function of the first kind.  If $\mu=0$, as for $x_i \sim \mathcal N(0,1)$, then $F_{1,1}=1$ and only the $\Gamma$ function contributes to the moments.

If $m<-1$,  $\vev{|y|^m}$ may diverge if $P(y=0)$ does not vanish fast enough. This is indeed the case for $x_i \sim \mathcal N(0,1)$, and we cannot predict the distribution of the sums in Eq.~\eqref{eqn:sums}  with $m\le-1$.    Sums like Eq.~\eqref{eqn:sums} are effectively finite numerical approximations to the integral
\begin{equation}
  \frac{1}{\Nf} \sum_i \lambda_i^n | x_i |^m \approx \int  |x|^m \mathcal N(0,1) dx  \int \lambda^n P(\lambda) d\lambda,
  \label{eqn:XXX}
\end{equation}
which diverges for $m<-1$.  While ratios of these sums might be well-defined~\cite{Frazer:2013zoa}, our approach shows that a finite prediction for both the mean and the standard deviation of $n_t/r$ requires $p>3/4$, while only requiring a finite mean needs $p>1/2$, using the CLT.

{\bf The method---}
Since $n_t/r$ is given by Eq.~\eqref{eqn:ntr_gen} and the sums in Eq.~\eqref{eqn:sums}  are ratios of correlated, normally distributed RVs, the key tool for this analysis is the ratio distribution $f_\mathrm{ratio}(\alpha/\beta)$ for normally distributed RVs $\alpha$ and $\beta$.  If $w \equiv \alpha/\beta$, then as $P(\beta>0) \to 1$ the CDF for the ratio distribution $f_\mathrm{ratio}(w)$ is approximately \cite{hinkley1969ratio}
\begin{equation}
  F_\mathrm{ratio}(w) = \Phi \left[ \frac{ \mu_\beta w - \mu_\alpha}{\sigma_\alpha \sigma_\beta a(w)} \right],
  \label{eqn:ratioCDF}
\end{equation}
where $\mu_i$ and $\sigma_i^2$ are the respective means and variances,
\begin{equation}
  a(w) \equiv \left[ \frac{w^2}{\sigma_\alpha^2} - \frac{2 \gamma w}{\sigma_\alpha \sigma_\beta} + \frac{1}{\sigma_\beta^2} \right]^{1/2},
  \label{eqn:XXX}
\end{equation}
and
\begin{equation}
  \Phi (z) \equiv \frac{1}{2} \left[ 1 + \mathrm{Erf}\left(\frac{z}{\sqrt{2}} \right) \right]
  \label{eqn:erf}
\end{equation}
for real $z$.  When $\Nf$ is large, $f_{\mathrm{ratio}}$ approaches a normal distribution with mean $\mu_\alpha/\mu_\beta$ and standard deviation
\begin{equation}
  s = \frac{\sqrt{\mu_\beta^2 \sigma_\alpha^2 - 2 \gamma \mu_\alpha \mu_\beta \sigma_\alpha \sigma_\beta + \mu_\alpha^2 \sigma_\beta^2}}{\mu_\beta^2}.
  \label{eqn:std_many}
\end{equation}
The mean of $f_\mathrm{ratio}$ is independent of the correlations $\gamma$, and the standard deviation for $n_t/r$ is a straightforward --- but messy --- algebraic function of $\vev{\lambda}$, $\vev{\lambda^2}$, and $\vev{\lambda^4}$, as well as $\vev{|x|^m}$ for $m=2,4,p,2p,2p-2$, and $4p-4$.

To obtain the distribution $f_\mathrm{ratio}(n_t/r)$ we express the consistency relation in terms of the sums in Eq.~\eqref{eqn:sums} as
\begin{equation}
\frac{n_t}{r} = -\frac{p N_*}{4} \left[ \frac{\sum_i \lambda_i^2 | \phi_{i,*}|^{2p-2}}{\left(\sum_j \lambda_j | \phi_j|^p \right)^2} \right].
  \label{eqn:ntr_expand}
\end{equation}
For each sum above, we  calculate the covariance in Eq.~\eqref{eqn:sums} between the numerator and denominator given $P(\lambda)$, and  use Eq.~\eqref{eqn:std_many} to find the variance of the sum.  Although the denominator $(\sum_i \lambda_i |\phi_{i,*}|^p)^2$ is then $\chi^2$-distributed, this is approximately normal in the many-field limit.  We  then substitute these two normally-distributed RVs back into Eq.~\eqref{eqn:ratioCDF}.  Similarly, we  evaluate the  correlation between the numerator and denominator in Eq.~\eqref{eqn:ntr_expand}, finally obtaining the  probability distribution for $n_t/r$.

\begin{figure}
  \includegraphics{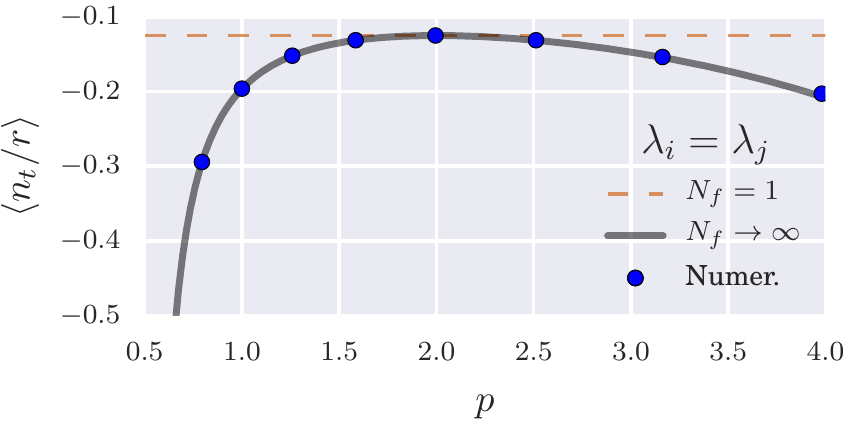}
  \caption{The multifield prediction from Eq.~\eqref{eqn:ntr_lln} compared to the numerical mean $\vev{n_t/r}$ of simulations with 5000 samples, at each plotted value of $p$, with $\Nf=1000$ using the horizon-crossing approximation.  The field values $\phi_{i,*}$ as the pivot scale $k_*$ leaves the horizon are drawn from a uniform prior on the surface in Eq.~\eqref{eqn:N_hca} and all the couplings $\lambda_i$ are identical. }
  \label{fig:hca_fixed_lambda}
\end{figure}

{ \bf Novel multifield predictions--- }
From the ratio distribution~\eqref{eqn:ratioCDF}, as $\Nf \to \infty$ the value of $n_t/r$ in Eq.~\eqref{eqn:ntr_expand} is normally distributed with a mean
\begin{equation}
  \label{eqn:ntr_lln}
  \vev{ \frac{n_t}{r} }_{\Nf \uparrow} = \left[- \frac{1}{8} \right] \left[ \frac{\vev{\lambda^2}}{\vev{\lambda}^2} \right] \left[ \frac{\sqrt{\pi} \; \Gamma \left(p - \frac{1}{2} \right)}{2 \; \Gamma^2 \left(\frac{p+1}{2} \right)} \right]
\end{equation}
and a standard deviation proportional to
\begin{equation}
  s_{n_t/r} \propto \frac{1}{\sqrt{\Nf}} \to 0 \quad \mathrm{as} \quad \Nf \to \infty,
  \label{eqn:standard}
\end{equation}
which can be found by substituting the means, variances, and correlations of Eq.~\eqref{eqn:sums} into Eq.~\eqref{eqn:std_many}.

The first bracketed term in Eq.~\eqref{eqn:ntr_lln} is the single-field prediction, the second is due to the couplings $\lambda_i$, and the third arises from the uniform prior for $\phi_{i,*}$ on the horizon-crossing surface.  This last term is due only to the spread in the field values at horizon crossing and is independent of everything except $p$.
The functional form of this term is fixed by the uniform prior distribution on the horizon crossing surface, but other prior probabilities for $\phi_{i,*}$ result in qualitatively similar behavior as demonstrated in Ref.~\cite{Easther:2013rva}.
As Eq.~\eqref{eqn:standard} vanishes in the many-field limit, Eq.~\eqref{eqn:ntr_lln} is the generic multifield prediction, which deviates from the single-field result at $>5 \sigma$ for $\Nf \gtrsim \order{10^2}$ for typical $P(\lambda)$.

Consequently, even if $\vev{\lambda^2}=\vev{\lambda}^2$,  $\Nf$-monomial models do not predict $n_t/r=-1/8$, unless the $\phi_{i,*}$ are also identical.
Fig.~\ref{fig:hca_fixed_lambda} compares the predicted value for $\vev{n_t/r}$ in Eq.~\eqref{eqn:ntr_lln}, with all $\lambda_i $ equal, to numerical results obtained by directly evaluating $n_t/r$ with Eq.~\eqref{eqn:ntr_gen}, showing excellent agreement for many fields.  The divergence at $p=1/2$ reflects the fact that $\vev{|x|^{2p-2}} \to \infty$.  Thus, when $p\le 1/2$,  $\vev{n_t/r}$ may be arbitrarily large, which violates the slow-roll assumption.  Consequently, these models are most easily distinguished from their single field analogues, but the hardest to make accurate predictions for.

\begin{figure}
  \includegraphics{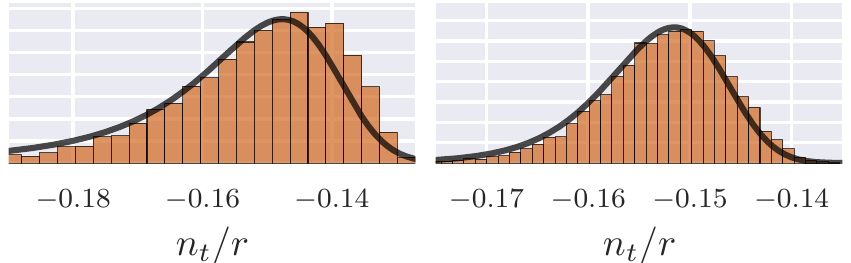}
  \caption{Predicted probability distributions for $n_t/r$ with $p=2$ compared with histograms built from 10000 numerical samples.  The couplings $\lambda_i$ are drawn from the Uniform Model with (\emph{left}) $\Nf=20$ and (\emph{right}) $\Nf=100$.  For $\Nf \lesssim \mathcal O(10^2)$, the distribution is skewed toward positive values as predicted.}
  \label{fig:hca_pred}
\end{figure}

{\bf Specific examples---}
To understand how the mean $\vev{n_t/r}$ in Eq.~\eqref{eqn:ntr_lln} is affected by $P(\lambda)$ we compare two explicit priors that are widely used in Bayesian analyses of inflation~\cite{Mortonson:2010er,2011PhRvD..83f3524M,Easther:2011yq,Ade:2013uln,Martin:2013nzq}.  We focus on the $p=2$ case, since the dependence on the prior on $\phi_{i,*}$ in Eq.~\eqref{eqn:ntr_lln} cancels for this scenario.

We look at two cases: uniform prior probabilities over $\lambda_i$ or $\alpha_i$ for $\lambda_i \equiv 10^{\alpha_i}$, which we denote the \emph{Uniform Model} and \emph{Log Model}, respectively. The Uniform Model would be applicable when the relevant scale of $\lambda_i$ is known to within an order of magnitude, while the Log Model effectively scans over a range of physical scales.

For the Uniform Model, the $\lambda_i$ are drawn from $\mathcal U[a,b]$, and Eq.~\eqref{eqn:ntr_lln} becomes
\begin{equation}
  \left(\frac{n_t}{r} \right)_{p=2}^\mathrm{unif} = -\frac{1}{6} \left[ \frac{b^2 +ab + a^2}{(b+a)^2} \right].
  \label{eqn:ntr_unif}
\end{equation}
For $ \lambda_i \in [10^{-14},10^{-13}]$, as $\Nf \to \infty$ the predicted correlation coefficient for $f_\mathrm{ratio}(n_t/r)$ is $\gamma \approx -0.98$ and $\vev{n_t/r} = -0.153$.  We plot $f_\mathrm{ratio}$ and the results of 10000 numerical realizations using the HCA in Fig.~\ref{fig:hca_pred}.  We find excellent agreement with Eq.~\eqref{eqn:ntr_unif}, with $f_\mathrm{ratio}$  accurately capturing the higher order moments of the $n_t/r$ distribution for $\Nf \gtrsim 20$.  For $p=\{3/2, 2, 3\}$ the single-field result $n_t/r=-1/8$ is a 5$\sigma$ deviation from the mean in Eq.~\eqref{eqn:ntr_unif} for $\Nf \gtrsim \{120, 120, 200\}$, respectively.

For the Log Model  with  $\alpha \sim \mathcal U[a,b]$,
\begin{equation}
  \left(\frac{n_t}{r} \right)_{p=2}^\mathrm{log} = - \frac{\log(10) (b-a)}{16 } \left[ \frac{10^b+10^a}{10^b-10^a} \right].
  \label{eqn:ntr_log}
\end{equation}
If $a \to b$, we recover the single-field result in both Eqs.~\eqref{eqn:ntr_unif}~and~\eqref{eqn:ntr_log}.  However, Eq.~\eqref{eqn:ntr_log} diverges as $a \to -\infty$, reflecting the failure of slow-roll in the limit of widely separated scales.  For $\alpha \in [-14,-12]$  the Log Model predicts $\mathcal P_\zeta \sim \mathcal O(10^{-9})$, $\epsilon \lesssim 0.03$, $\gamma \approx -0.95$ and $n_t/r = -0.294$.   For $p=\{3/2, 2, 3\}$ the single-field result is a 5$\sigma$ deviation from the mean in Eq.~\eqref{eqn:ntr_log} for $\Nf \gtrsim \{145, 135, 255\}$, respectively.

\begin{figure}
  \includegraphics{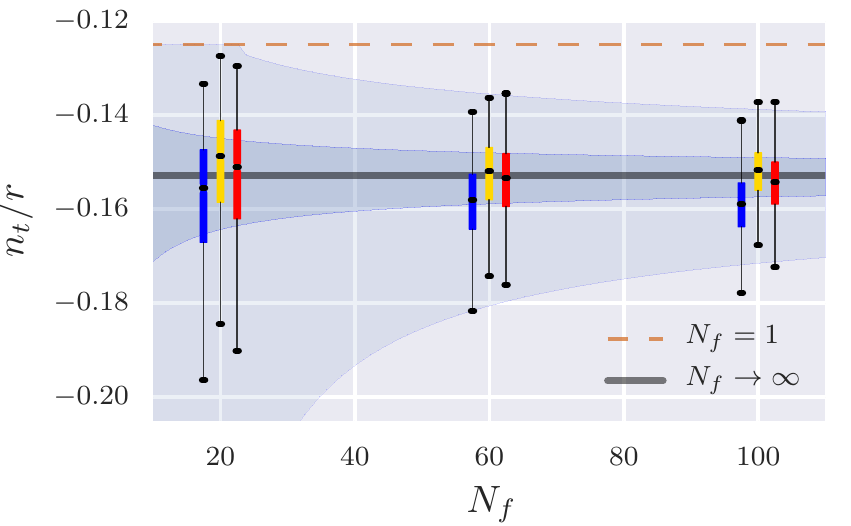}
  \caption{The consistency relation for the Uniform Model with $p=2$ is plotted for different $\Nf$, marginalizing over initial field values.  The boxes/whiskers cover the $50/97\%$ CIs and the gray regions delineate the same ranges as predicted by the HCA and the central limit theorem. The (\emph{dashed}) brown and (\emph{solid}) gray lines are the single-field and the many-field HCA predictions, respectively.  For each case we present  results derived from full numerical solutions to the mode equations (\emph{blue/left}), the slow-roll prediction using the HCA (\emph{yellow/center}), and  the slow-roll prediction including the end-of-inflation surface (\emph{red/right}) for $\Nf=20$, $60$, and $100$.}
  \label{fig:compare}
\end{figure}

{\bf Relaxing the approximations---}
Fig.~\ref{fig:compare} compares the HCA prediction to numerical results that include the contribution from the end-of-inflation surface in Eq.~\eqref{eqn:N_hca}, with $\phi_{i,c} \ne 0$.  We  numerically solve the background Klein-Gordon equations for 1000 realizations, finding the field values at the end of inflation (defined by $\epsilon =1$) and obtaining the full $\delta N$ prediction without using the HCA.  Fig.~\ref{fig:compare} also incorporates both the sub-horizon evolution of the modes and any non--slow-roll behavior by solving the mode equations numerically, as in Refs.~\cite{Salopek:1988qh, Easther:2013rva}, using {\sc MultiModeCode}~\cite{Price:2014xpa}.  Results are plotted for the Uniform Model, with the ranges $ \lambda_i \in [10^{-14},10^{-13}]$ and $p=2$.

In all cases the numerical results are well-approximated by the HCA. The HCA results are marginally larger than the numerical results, which we attribute to second-order corrections to the slow-roll equations;  $n_t = -2 \epsilon/(1-\epsilon)$, which suppresses $n_t$ relative to the first-order approximation.
The  variances in the numerical  results scale as $\sigma^2 \propto 1/\sqrt{\Nf}$, as predicted by the HCA results, confirming that many-field models make sharp predictions for $n_t/r$.

{\bf Conclusion---}
We have computed the probability distribution for the consistency relation $n_t/r$ for inflation driven by multiple scalar fields with monomial potential terms, as a function of the distribution of couplings and initial field values.  The single-field result is clearly distinguishable from the many-field limit, providing a clean and compelling signature that distinguishes these models from their single-field analogues.
Other than for the quadratic case, this result holds even when the couplings are identical.

We focused on computing the slow roll parameter $\epsilon$, but the  nature of the slow-roll hierarchy~\cite{Easther:2005nh} indicates that this approach will generalize to a variety of observables, so quantities such as $f_\mathrm{NL}$ that rely on the second and higher slow-roll parameters should also have precise predictions that deviate from the single-field expectation even when the couplings are degenerate.
This provides a further compelling example of a multifield scenario in which the observables have a sharp and  \emph{generic} prediction in the many-field limit \cite{Aazami:2005jf,Alabidi:2005qi,Easther:2005zr,Kim:2006ys,Kim:2006te,Piao:2006nm,Kim:2007bc,Frazer:2013zoa,Kaiser:2013sna,Kallosh:2013hoa,Kallosh:2013maa,Kallosh:2013daa,Easther:2013rva,Sloan:2014jra}.

The expected value $\vev{n_t/r}$ depends on only two moments of the prior probability distributions $P(\lambda)$ and $P(\phi_0)$, and the corresponding variance is $s_{n_t/r}^2 \propto 1/\Nf$.   The single-field expectation of $n_t/r=-1/8$ differs from the multifield result at the $5\sigma$ level when $\Nf \gtrsim \mathcal O(10^2)$.  Consequently, given specific priors for the field values and couplings, we obtain generic and testable predictions for the consistency relations in this large and interesting class of multifield inflation models.

\acknowledgments

\paragraph{Acknowledgments---}
We thank Grigor Aslanyan, Andrew Jaffe, and Jonathan White for helpful discussions.  HVP is supported by STFC and the European Research Council under the European Community's Seventh Framework Programme (FP7/2007-2013) / ERC grant agreement no 306478-CosmicDawn. JF is supported by IKERBASQUE, the Basque Foundation for Science.  We acknowledge the contribution of the NeSI high-performance computing facilities. New Zealand's national facilities are provided by the New Zealand eScience Infrastructure (NeSI) and funded jointly by NeSI's collaborator institutions and through the Ministry of Business, Innovation \& Employment's Research Infrastructure programme \footnote{{\url{http://www.nesi.org.nz}}}. This work has been facilitated by the Royal Society under their International Exchanges Scheme. This work was supported in part by National Science Foundation Grant No. PHYS-1066293 and the hospitality of the Aspen Center for Physics.

\bibliographystyle{apsrev4-1}
\bibliography{references}

\end{document}